\documentclass[aps,prd,amssymb,amsmath,reprint]{revtex4-1}
\begin{document}
\title{Introduction to Quantum Gravity}
\author{Ricardo Paszko}
\email{ricardo.paszko@ufabc.edu.br}
\affiliation{Centro de Ci\^{e}ncias Naturais e Humanas, Universidade Federal do ABC, 09210-170, Santo Andr\'{e}, SP, Brazil}
\begin{abstract}
In this talk\footnote{Invited talk held at UNESP, Guaratinguet\'{a}, 11/23/2011.}, we give a glimpse of the problems with quantum gravity and some possible solutions.
\end{abstract}
\maketitle
\section{Introduction}	
Over the centuries, physicists try to unify the fundamental interactions. An essential ingredient for these unification theories is the gauge symmetry, in particular, the (special) unitary group.
\begin{center}
\begin{tabular}{lc}
Unification&Group\\
\hline
Electromagnetic&$U(1)$\\
Electroweak&$SU(2)\times U(1)$\\
Standard Model&$SU(3)\times SU(2)\times U(1)$\\
Grand Unified Theory&$?\times SU(3)\times SU(2)\times U(1)$
\end{tabular}
\end{center}
However, among the $4$ fundamental interactions in nature, only the gravitational interaction is related to other group than the unitary. Furthermore, it is described by other lagrangian than the Yang-Mills (despite the ``accident'' in $(2+1)$-dimensions \cite{Witten}).
\begin{center}
\begin{tabular}{lcl}
Interaction&Group&Lagrangian\\
\hline
Strong&$SU(3)$&\vspace{-.29cm}\\
Weak&$SU(2)$&$\Bigg\}$Yang-Mills\vspace{-.29cm}\\
Electromagnetic&$U(1)$&\\
Gravitational&General Covariance&Einstein-Hilbert
\end{tabular}
\end{center}
Therefore, from the unification viewpoint, it would be reasonable to argue that general covariance is equivalent to some unitary group and/or is the Einstein-Hilbert lagrangian tantamount to Yang-Mills lagrangian? As we will see, the answer to both questions is no.
\section{Comparison between Yang-Mills and Einstein-Hilbert lagrangians}
The Yang-Mills lagrangian
\[{\cal L}_{YM}=\frac{1}{\kappa^2}{\rm tr}(\star F\wedge F),\]
where the curvature $F=dA+A\wedge A$, $\star$ is the Hodge dual operator, ${\rm tr}$ is the trace over $\mathfrak{su}(N)$ algebra and $\kappa$ is the coupling constant, can be rewritten, after rescaling the gauge field $A\to\kappa A$, symbolically as
\begin{equation}
{\cal L}_{YM}=\underbrace{(dA)^2}_{{\cal L}_{\rm free}}+\underbrace{\kappa(dA)A^2+\kappa^2A^4}_{{\cal L}_{\rm interaction}},\label{polynomial}
\end{equation}
and thus is easy to quantize --- a free propagator and a finite number of (two) vertices --- at least in the high-energy limit, where $\kappa$ is small and perturbation theory makes sense.

On the other hand, the Einstein-Hilbert lagrangian
\[{\cal L}_{EH}=\frac{1}{\kappa^2}\sqrt{g}R,\]
where $\kappa^2=16\pi G$, $G$ is Newton's constant, the scalar curvature $R=g^{\mu\nu}(\Gamma^\alpha_{\mu\alpha,\nu}-\Gamma^\alpha_{\mu\nu,\alpha}+\Gamma^\beta_{\mu\alpha}\Gamma^\alpha_{\beta\nu}-\Gamma^\beta_{\mu\nu}\Gamma^\alpha_{\beta\alpha})$, the Christoffel symbol $\Gamma^\alpha_{\mu\nu}=\frac{1}{2}g^{\alpha\beta}(g_{\nu\beta,\mu}+g_{\mu\beta,\nu}-g_{\mu\nu,\beta})$ and $g=\det(g_{\mu\nu})$ is the determinant of the metric $g_{\mu\nu}$, is distinct from the Yang-Mills lagrangian. Even the equivalent $\Gamma\Gamma$ lagrangian \cite{Dirac}
\[{\cal L}_{\Gamma\Gamma}=\frac{1}{\kappa^2}\sqrt{g}g^{\mu\nu}(\Gamma^\beta_{\mu\nu}\Gamma^\alpha_{\beta\alpha}-\Gamma^\beta_{\mu\alpha}\Gamma^\alpha_{\beta\nu})\]
is completely different from Yang-Mills, as can be seen explicitly in some $D$-dimensional examples
\begin{align}
D=1,\qquad&{\cal L}_{\Gamma\Gamma}=0\nonumber\\
D=2,\qquad&{\cal L}_{\Gamma\Gamma}=\frac{1}{2\kappa^2}\frac{
\begin{vmatrix}
g_{11}&g_{12}&g_{22}\\
g_{11,1}&g_{12,1}&g_{22,1}\\
g_{11,2}&g_{12,2}&g_{22,2}
\end{vmatrix}}
{\begin{vmatrix}
g_{11}&g_{12}\\
g_{12}&g_{22}
\end{vmatrix}^{3/2}}\label{determinants}\\
D=3,\qquad&{\cal L}_{\Gamma\Gamma}=\cdots\nonumber
\end{align}
(after some algebra, the $2$-dimensional case simplifies to this fraction of determinants, this simplification doesn't seem to occur for higher dimensions). Proceeding as before, by rescaling the metric $g_{\mu\nu}\to\kappa g_{\mu\nu}$, is useless, specially in the $2$-dimensional case which is scale invariant. So, how to quantize?
\section{The Quantization Problem}
\subsection{Quantization methods}
There exist several methods of quantization: canonical, constrained, path-integral, stochastic, etc. There are pros and cons to each method, but all are solvable, roughly, only for quadratic terms in the lagrangian. For example, the path-integral of the free Yang-Mills lagrangian, Eq.~(\ref{polynomial}) with $\kappa=0$, plus a source term $J(x)$,
\begin{align*}
Z_{\kappa=0}[J]&=\int DA\,e^{-\int d^4x({\cal L}_{free}+J\cdot A)}\\
&=\int DA\,e^{-\int d^4x[(dA)^2+J\cdot A]},
\end{align*}
can be done exactly, since these are gaussian integrals. From the functional generator $Z[J]$, through differentiation in $J$, the propagator, vertices, etc. can be calculated. Some details still need attention, such as ghosts, owing to gauge invariance, although the remaining is just a power series expansion in $\kappa$. How about the free Einstein-Hilbert lagrangian? Eqs.~(\ref{determinants}) with $\kappa=0$?
\subsection{Early attempts}
Historically, the first attempt to quantize gravity was done by Rosenfeld \cite{Rosenfeld} through a ``scale-shift'' transformation, best known as a fixed background field, i.e.,
\[g_{\mu\nu}(x)=\delta_{\mu\nu}+\kappa h_{\mu\nu}(x),\]
(in this case the expansion is around a flat spacetime, where $g^{\kappa=0}_{\mu\nu}(x)=\delta_{\mu\nu}$ is the euclidean metric) such that, after expansion in $\kappa$, the Einstein-Hilbert or the $\Gamma\Gamma$ lagrangian become
\[{\cal L}_{EH}=\underbrace{(d h)^2}_{{\cal L}_{free}}+\underbrace{\kappa(d h)^2h+\kappa^2(d h)^2h^2+\kappa^3(d h)^2h^3+\cdots}_{{\cal L}_{interaction}}\]
which resembles the Yang-Mills lagrangian, Eq.~(\ref{polynomial}), but has an infinite number of vertices.

Other attempts employing different types of transformations, for example, another choices of scale-shift such as $\sqrt{g}g_{\mu\nu}=\delta_{\mu\nu}+\kappa h_{\mu\nu}$, $g^{\mu\nu}=\delta^{\mu\nu}-\kappa h^{\mu\nu}$, $\ldots$ Or distinct background fields $g_{\mu\nu}=\varphi_{\mu\nu}+\kappa h_{\mu\nu}$, where $\varphi_{\mu\nu}$ is, e.g., the (anti) de Sitter metric. Still another type using the first order formalism as $e_\mu^i=\delta_\mu^i+\kappa h_\mu^i$, where $e_\mu^i$ is the tetrad field, etc. All have an identical problem --- the renormalization problem --- after all, they share the same properties, i.e., a fixed background field and a perturbative expansion in a dimensional coupling constant $\kappa\propto\sqrt{G}$ with length dimension \cite{Weinberg}.
\subsection{The renormalization problem}
Despite the initial miracle \cite{tHooft} in the pure gravitational sector, in the presence of matter fields the one-loop calculation has a divergence of the form $c_1R^2+c_2R^{\mu\nu}R_{\mu\nu}$, where $c_1$ and $c_2$ are constants. One possible way out was to modify the Einstein-Hilbert lagrangian to a higher-derivative lagrangian
\[{\cal L}=\frac{1}{\kappa^2}\sqrt{g}(R+\alpha R^2+\beta R^{\mu\nu}R_{\mu\nu}),\]
which makes the theory renormalizable \cite{Stelle} but, unfortunately, suffers from other problems \cite{Antoniadis}. Another way out is to consider a more general function in the lagrangian $f(R)=R+\alpha R^2+\beta R^{\mu\nu}R_{\mu\nu}+\cdots$, although it becomes more difficult the interpretation of the constants $\alpha,\,\beta,\,\ldots$ because of the lack of experimental data, since the energy scale is the Planck energy $1/\kappa\propto1/\sqrt{G}\sim10^{28}$eV.
\section{Some Recent Theories}
\subsection{Perturbative theories}
The renormalization problem found using perturbation theory has alternative solutions, to wit: the semiclassical \cite{Moller} and effective \cite{Donoghue} methods. In the former, gravity is a classical field and everything else is quantized; now in the latter, everything is quantized, including gravity, but the Feynman amplitude is expanded in terms of the momentum exchanged. Despite being quite different approaches, they are equivalent \cite{Paszko} in the sense that both give results strictly valid in the low-energy limit.
\subsection{Non-perturbative theories}
We do not intend to discuss these theories, although some comments are in order. As far as we know, even in the $2$-dimensional toy model, there is no choice of variables that makes the fraction of determinants, Eq.~(\ref{determinants}), into a fourth order polynomial, Eq.~(\ref{polynomial}), at least without ruining the measure in the path-integral as in the unimodular gravity \cite{unimodular}. Therefore, other choices of variables seem unhelpful, specially in the $4$-dimensional case. Besides that, theories of everything have struggled to overcome worse problems than the renormalization problem. In view of Occam's razor, it is reasonable to search for a simpler solution.
\subsection{Modified BF theory}
A possible solution to the presented problems is a expansion in the dimensionless and extremely small coupling constant $\kappa^2=G\Lambda\sim10^{-120}$, where $\Lambda$ is the cosmological constant. This is achieved through a modification of the BF theory lagrangian ${\cal L}={\rm tr}(-iB\wedge F)$, namely,
\begin{equation}
{\cal L}={\rm tr}(-iB\wedge F-\kappa^2B\wedge\Gamma B),
\end{equation}
which is polynomial, and for $\Gamma=\gamma_5$ reproduces the Einstein-Hilbert lagrangian \cite{Smolin}, and for $\Gamma=\star$ also reproduces the Yang-Mills theory \cite{Cattaneo}.

In this approach, the metric is a derived object, i.e., the $SO(5)$ gauge field $A_\mu^{IJ}$ breaks into a $SO(4)$ spin connection $A_\mu^{ij}=\omega_\mu^{ij}$ and a tetrad field $A_\mu^{i5}=\sqrt{\Lambda}e_\mu^i$, and thus
\begin{equation}
g_{\mu\nu}=e_\mu^ie_\nu^j\delta_{ij}=\frac{1}{\Lambda}A_\mu^{i5}A_\nu^{j5}\delta_{ij}.\label{square}
\end{equation}
Even in the lorentzian case the gauge group is $SO(4,1)$ or $SO(3,2)$, and hence the general covariance group seems related to the (broken symmetry) special orthogonal group. Moreover, as an expansion around a topological theory, it is background independent.

Despite some discussions \cite{Rovelli}, common to new ideas, it is worth noting that Eq.~(\ref{square}) is a sign of the conjecture that gravity is the square of a gauge field \cite{Bern}.
\section{Conclusions}
In this talk, we give a glimpse of the quantization problem of gravity. As argued, this problem is due to the non-polynomial form of the Einstein-Hilbert lagrangian. Attempts to recast this expression into a polynomial form, as the fixed background field, seem unhelpful and lead to another problem --- the renormalization problem --- that is also related to the dimensional coupling constant $\kappa^2\propto G$.

Therefore, a simple theory that can be written in polynomial form, with background independence and a dimensionless coupling constant, is a serious candidate as a possible solution to these problems. As we have seen, a theory that fulfills all these prerequisites is the Modified BF. Furthermore, from the unification point of view, this theory is worthwhile, inasmuch as it reproduces the Einstein-Hilbert and the Yang-Mills lagrangians.

\end{document}